\documentclass{article}

\usepackage[english]{babel}
\usepackage{todonotes}
\usepackage{tikz}
\usetikzlibrary{patterns}
\usepackage{authblk}
\usepackage{pgfplots}
 \usepackage{fullpage}%for printing
\usepackage{subcaption}
\usepackage{amsthm}
\usepackage{thmtools}
\usepackage{amsmath,amssymb}
\usepackage{mathtools}
\usepackage{graphicx}
\usepackage[colorlinks=true, allcolors=blue]{hyperref}
\usepackage[noabbrev,nameinlink]{cleveref}

\DeclareMathOperator*{\argmin}{arg\,min}
\newtheorem{definition}{Definition}[section]
\newtheorem{theorem}[definition]{Theorem}

\newtheorem{conjecture}[definition]{Conjecture}

\title{Outperforming the Best 1D Low-Discrepancy Constructions with a Greedy Algorithm}
\author{Fran\c{c}ois Cl\'ement\thanks{francois.clement@lip6.fr}}
 \affil{Sorbonne Universit\'e, CNRS, LIP6, Paris, France}

\begin{document}

\maketitle

\begin{abstract}
    The design of uniformly spread sequences on $[0,1)$ has been extensively studied since the work of Weyl~\cite{Weyl} and van der Corput~\cite{vdC} in the early $20^{\text{th}}$ century. The current best sequences are based on the Kronecker sequence with golden ratio %defined by $(\{n\phi\})_{n \in \mathbb{N}}$ where $\phi$ is the golden ratio and $\{\}$ the fractional part, 
    and a permutation of the van der Corput sequence by Ostromoukhov~\cite{Ostro}. Despite extensive efforts, it is still unclear if it is possible to improve these constructions further. We show, using numerical experiments, that a radically different approach introduced by Kritzinger in~\cite{Kritz} seems to perform better than the existing methods. In particular, this construction is based on a \emph{greedy} approach, and yet outperforms very delicate number-theoretic constructions. Furthermore, we are also able to provide the first numerical results in dimensions 2 and 3, and show that the sequence remains highly regular in this new setting.
\end{abstract}

\section{Introduction}\label{sec:intro}
Uniformly distributed sequences rely on discrepancy theory to characterize their regularity. Discrepancy measures are a family of measures that quantify how well a point set or sequence approximates the uniform distribution over a given space, usually $[0,1)^d$. Among these measures, the arguably most important one is the $L_{\infty}$ star discrepancy, largely because of the Koksma-Hlawka inequality~\cite{Koksma,Hlawka} and the multitude of applications it is involved in, \cite{CauwetCDLRRTTU20, DickP10,SantnerDoE,GalFin,MatBuilder, Maaranen} to name a few. The $L_{\infty}$ star discrepancy of a finite point \emph{set} $P \subseteq [0,1)^d$ measures the worst absolute difference between the Lebesgue measure of a $d$-dimensional box $[0,q)$ anchored in $(0,\ldots,0)$ and the proportion $|P \cap [0,q)|/|P|$ of points that fall inside this box. More formally, it is defined by
\begin{equation}
d^*_{\infty}(P) := \sup_{q\in[0,1)^d} \left| \frac{|P \cap [0,q)|}{|P|} - \lambda(q) \right|,
\label{eq:disc_def}
\end{equation}
where $\lambda(q)$ is the Lebesgue measure. The $L_{\infty}$ star discrepancy is always between 0 and 1, the lower it is, the more uniformly spread is the point set. For a finite point set in one dimension, Niederreiter gave the optimal construction in~\cite{NieBox}. For a set of $n$ points, it consists in $\{(2i+1)/(2n):i \in \{0,\ldots,n-1\}\}$ and has a discrepancy of $1/(2n)$. One can see that this corresponds to dividing the torus $\mathbb{T}$ of length 1 in $n$ intervals of equal length. 

More questions arise when we want to have an infinite \emph{sequence} that has as low discrepancy as possible. The $L_{\infty}$ star discrepancy of an infinite sequence $P'$ is given by the following function of $k$
\begin{equation}
    d^*_{\infty}(P',k) :=\sup_{q\in[0,1)^d} \left| \frac{|P'_k \cap [0,q)|}{k} - \lambda(q) \right|,
    \label{eq:dis_def_seq}
\end{equation} %In other words, the discrepancy is given by the worst discrepancy of any prefix of the sequence. 
where $P_{k}'$ corresponds to the first $k$ elements of $P'$. We are interested in how this function evolves with $k$. In other words, we are not focusing on the discrepancy of a single set but in the worst discrepancy of an increasing sequence of nested sets and how to bound it. In this setting, early results by van Aardenne-Ehrenfest~\cite{vAE} and Roth~\cite{Roth54} showed that the optimal order of discrepancy is $\Theta(\log(n)/n)$ for a sequence in one dimension. However, finding the optimal constant remains an open problem.

\begin{figure}[h]
    \centering
    \includegraphics[width=0.24\textwidth]{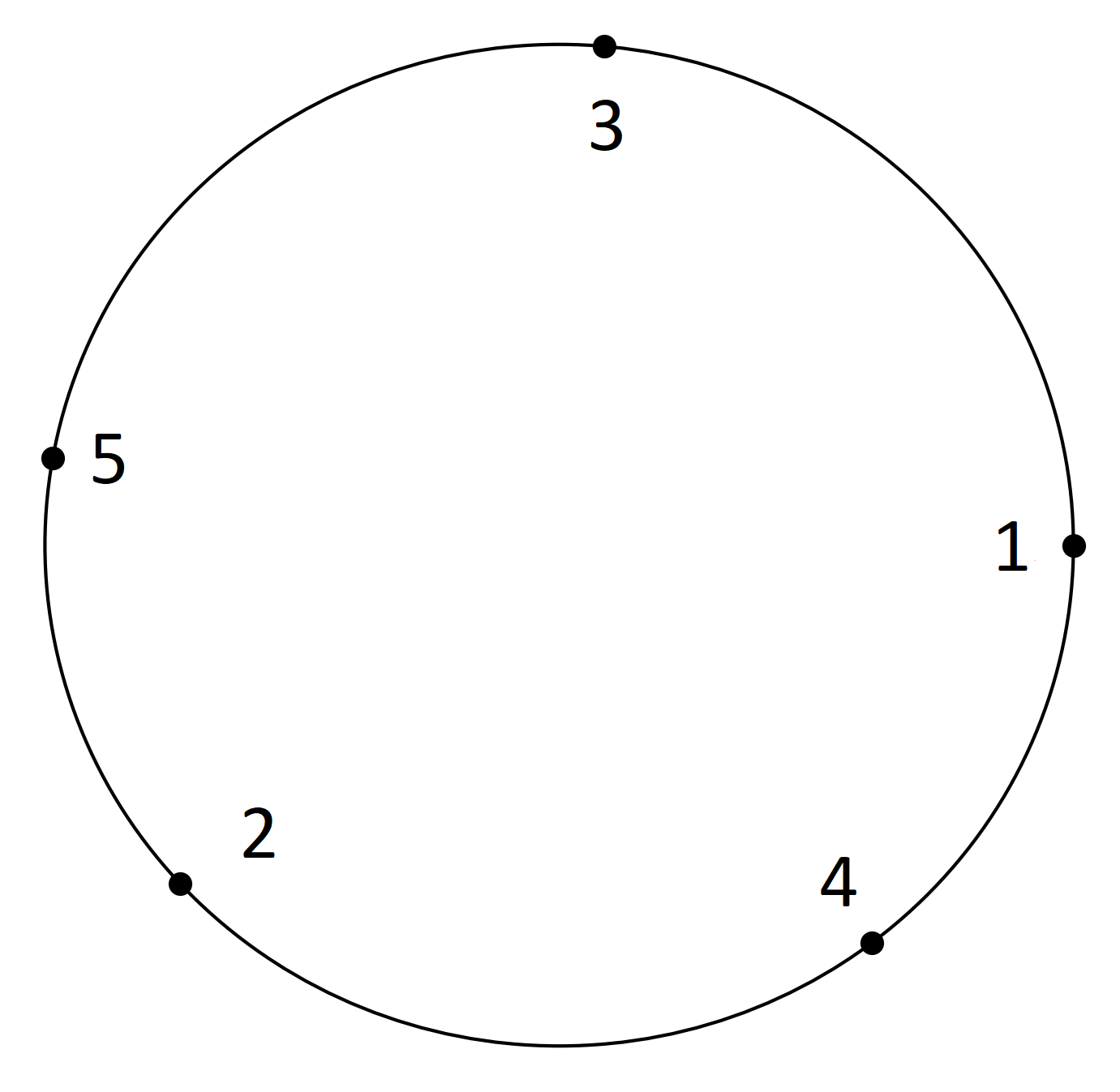}\hfill
    \includegraphics[width=0.24\textwidth]{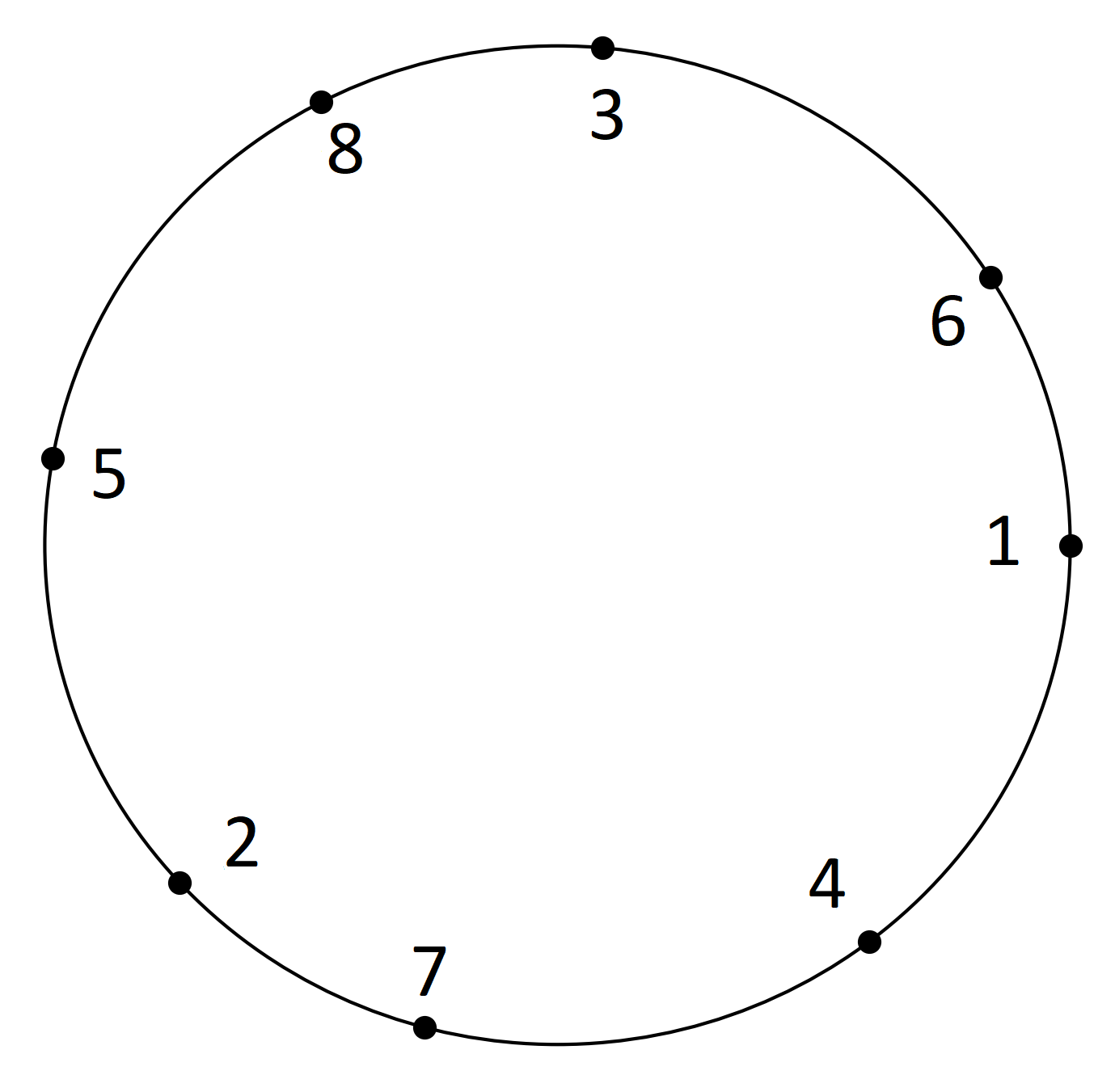}\hfill
    \includegraphics[width=0.24\textwidth]{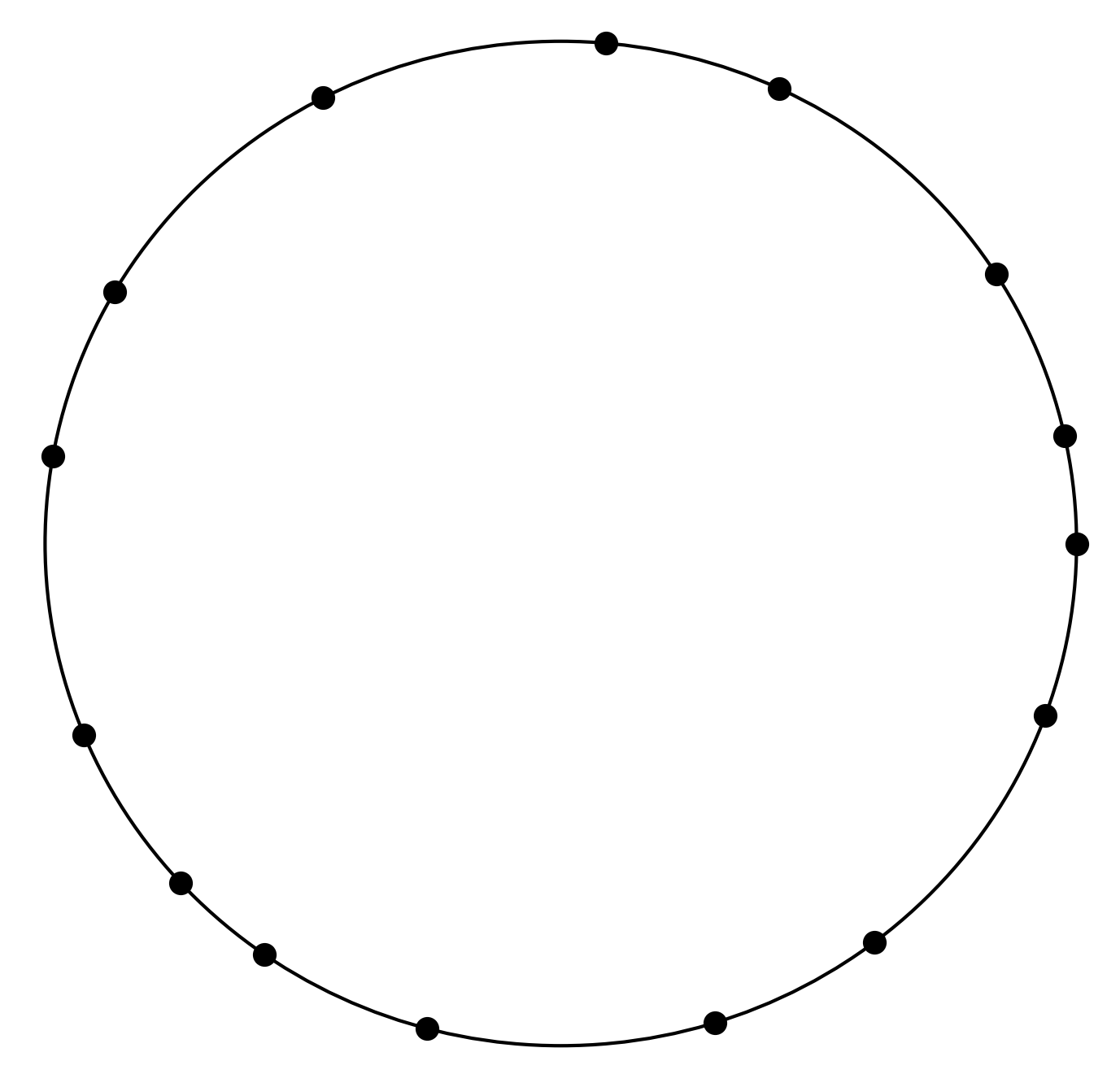}\hfill
    \includegraphics[width=0.24\textwidth]{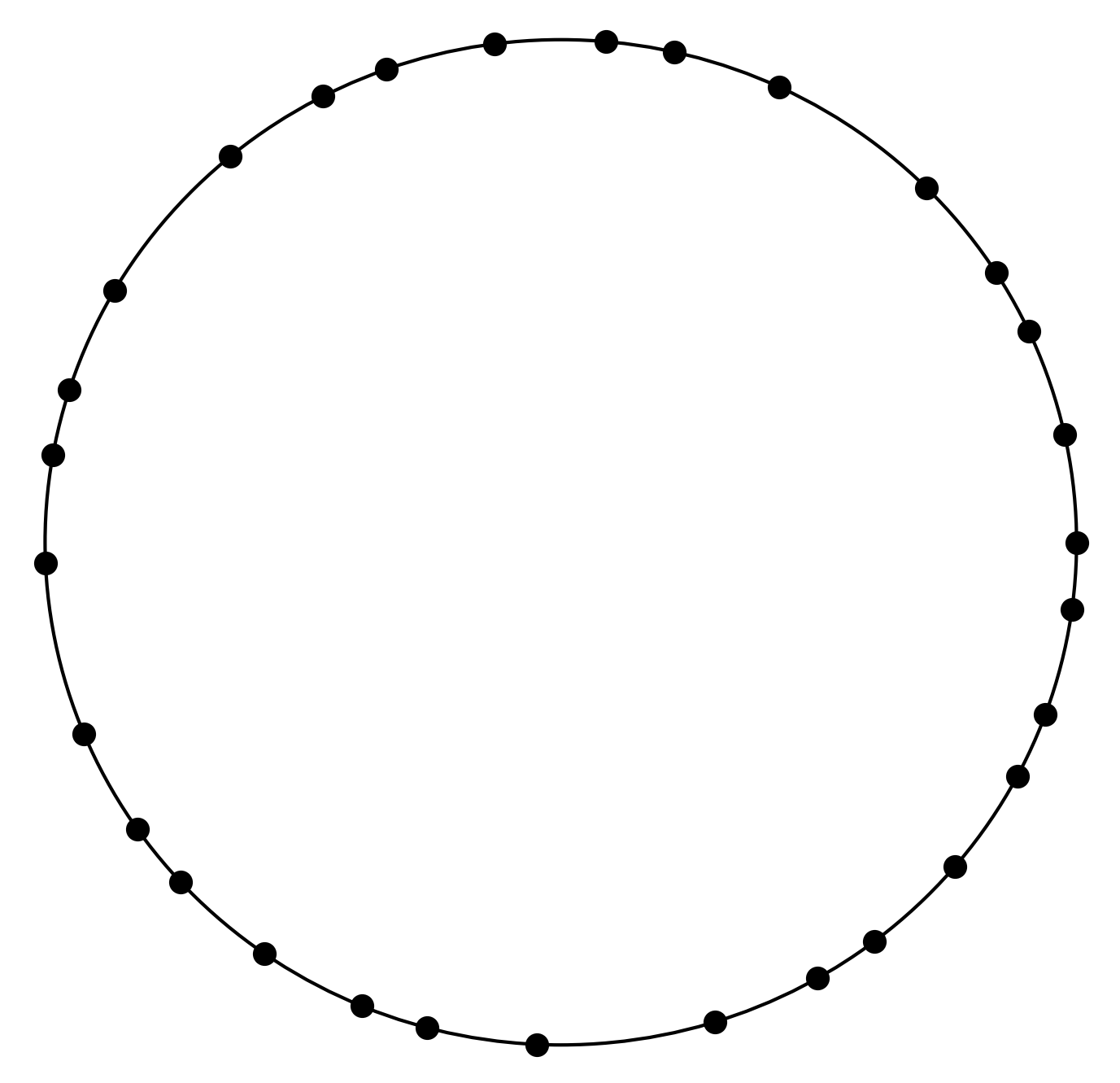}\hfill
    \caption{The Kronecker sequence with golden ratio for the first 5, 8, 15 and 30 elements, from left to right. We represent $[0,1)$ and the fractional part as a torus to better illustrate how the points are gradually more and more uniformly distributed. As shown in the two leftmost images by the numbering, the elements are added one by one starting from the first in (1,0), each time with a rotation of $2\phi\pi$ to add the next point.}
    \label{fig:Kron}
\end{figure}

There are two main families of constructions to obtain as low a constant as possible for the $L_{\infty}$ star discrepancy. The first corresponds to Kronecker sequences, of the shape $P_{\alpha}:=(\{n\alpha\}_{n \in \mathbb{N}})$, where $\alpha$ is irrational and $\{x\}$ is the fractional part of $x$. This is based on Weyl's criterion~\cite{Weyl}, which states that any such sequence is ``uniformly distributed'' for $\alpha$ irrational. Uniformly distributed does not give us the discrepancy convergence rate, but states that the discrepancy of the sequence converges to 0 when the number of points $n$ tends to infinity. \Cref{fig:Kron} describes how Kronecker sequences are built. The choice of $\alpha$ is very important for the regularity of the sequence, results described in~\cite{Nie92} show that the golden ratio $\phi=(1+\sqrt{5})/2$ should be the best choice. Empirically, it is the sequence with lowest $L_{\infty}$ star discrepancy known to this day (see~\cite{Nie92} for a bound on its discrepancy based on continuous fraction expansions).

\begin{figure}[h]
    \centering
    \includegraphics[width=0.47\textwidth]{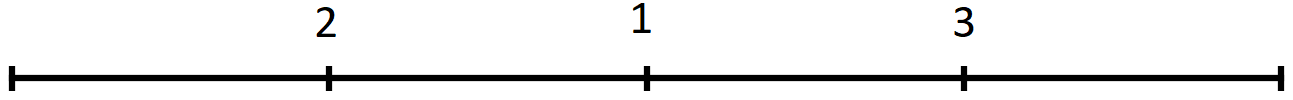}\hfill
    \includegraphics[width=0.47\textwidth]{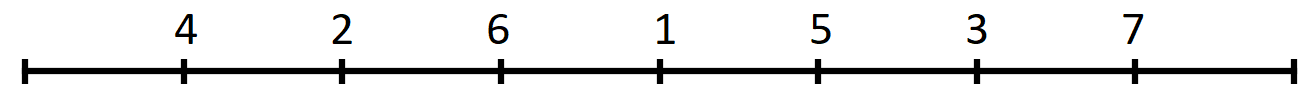}
    \caption{For both images, the segment corresponds to $[0,1]$. On the left, the first three elements of the van der Corput sequence. They correspond to the numbers that are written 01, 10 and 11 in binary. 01 gives $0.5$, 10 is $0.25$ while 11 corresponds to $0.75$. The next four elements to be added are exactly in the middle of the segments between the previous points. They are also added regularly: one element is added to each half of $[0,1]$ at first, then one to each quarter and so on.}
    \label{fig:vdC}
\end{figure}

However, the Kronecker sequence with golden ratio does not hold the record for the best theoretical asymptotic constant in the $\Theta(\log(n)/n)$ for the $L_{\infty}$ star discrepancy. The best constant is obtained with a permutation of the van der Corput sequence~\cite{vdC}. Permutations of this sequence are the second main type of one-dimensional low discrepancy constructions. The original van der Corput sequence is obtained by bit-reversing integers in base 2. Given an integer $i=\sum_{j \in \mathbb{N}}a_j2^{j}$, where $a_i \in \{0,1\}$, the $i$-th element of the sequence is given by $x_i:=\sum_{j \in \mathbb{N}}a_j2^{-j-1}$. \Cref{fig:vdC} describes this sequence, and shows that when adding the next element, it is added exactly halfway between the two most distant points on $[0,1)$. This guarantees that whenever we have a number of points equal to a power of two, the points are evenly spread and equal to the optimal set by Niederreiter.

This sequence can be further generalized by allowing a different base $b$ other than 2, and by allowing permutations of the $a_j$. Given an integer $i=\sum_{j \in \mathbb{N}}a_jb^{j}$, the $i$-th element of the generalized van der Corput sequence in base $b$ is given by $x_i:=\sum_{j \in \mathbb{N}}\pi_j(a_j)b^{-j-1}$, where the $\pi_j$ are permutations of $\{0,\ldots,b-1\}$. Using such a construction, Ostromoukhov obtained the best theoretical constant $0.2223$ in~\cite{Ostro} for the asymptotic $L_{\infty}$ star discrepancy of a one-dimensional sequence. However, there remains an important gap with the best known lower bound, $0.065$ by Larcher and Puchhammer~\cite{LarPuch}. %To summarize, past constructions are based either on the regularity of irrational rotations, or the regularity of digit expansions in prime bases.

In a bid to find new approaches to tackle this problem, Steinerberger proposed new \emph{greedy} methods of constructing point sets in \cite{StefEnerg,StefDyn}. The goal is to dynamically construct a sequence $(x_n)_{n\in \mathbb{N}}$, by defining the newest point in the sequence $x_{n+1}$ as
$$x_{n+1}=\argmin \sum_{1 \leq i \leq n} f(|x_{n+1}-x_i|), $$
where $f$ is an appropriately chosen function. While there is not yet a theoretical proof for the regularity of the resulting sequence for any $f$, these constructions appear to be very regular with a well-chosen function $f$. In particular, Pausinger showed in~\cite{Pausinger} that if $f$ verifies three simple conditions, then the resulting sequence is a permutation of the van der Corput sequence, and has an $L_{\infty}$ star discrepancy in $O(n^{-1/3})$. 

Based on these greedy approaches, Kritzinger introduced in~\cite{Kritz} a sequence construction using the following greedy choice (see~\Cref{sec:L2} for its origin). Given $P:=(x_i)_{i \in \{1,\ldots,n\}}$ a point set of size $n$, the next point $x_{n+1}$ is given by
$$x_{n+1}:=\argmin_{y \in [0,1)}\,(n+1)y^2-y-2\sum_{i=1}^{n}\max(x_i,y).$$

Kritzinger showed that regardless of the choice of the initial point(s), future points will always be rational and with only specific values possible, and that the resulting sequence has an $L_{\infty}$ star discrepancy in $O(n^{-1/2})$. Despite this clearly not being as good as the optimal $\Theta(\log(n)/n)$, his numerical experiments for up to around $1\,500$ points suggest that it is competitive with both the Kronecker sequence with golden ratio and the van der Corput sequence. Steinerberger later showed a bound of $O(n^{-2/3})$ for infinitely many $n$ in~\cite{StefL2}, complemented by numerical experiments with the same order of points suggesting it performed much better than the bound suggests.

\begin{figure}[h]
    \centering
    \includegraphics[width=0.47\textwidth]{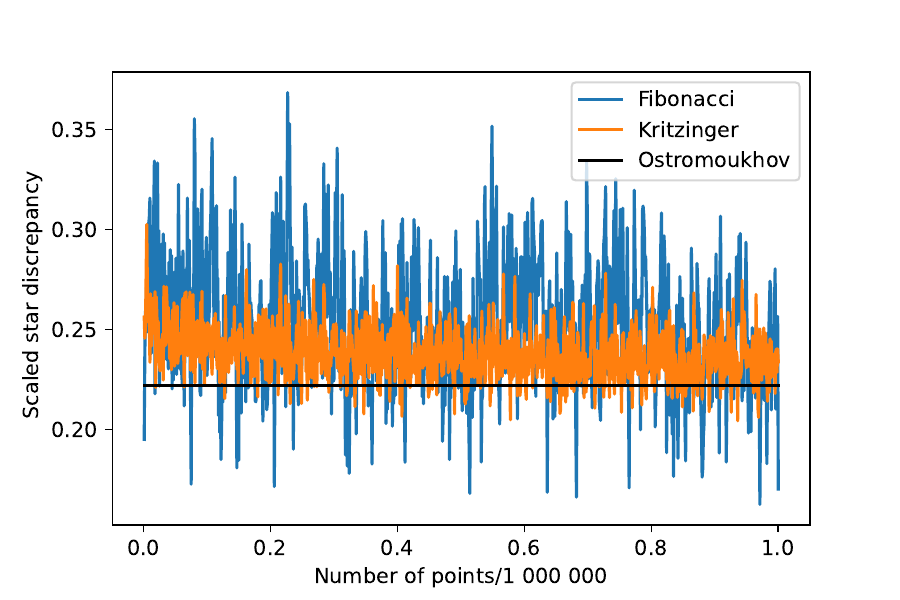}
    \includegraphics[width=0.47\textwidth]{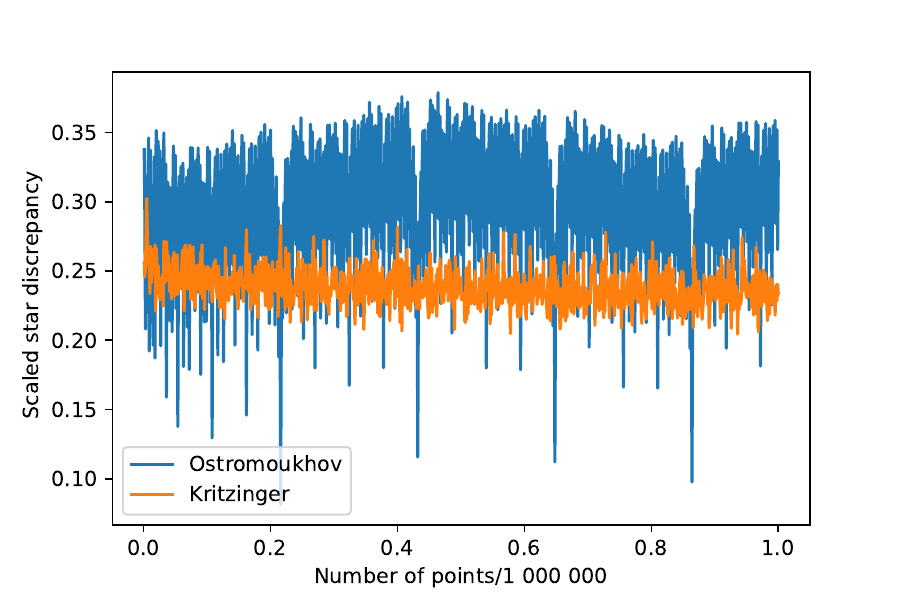}
    \caption{Comparison of the first million points of the Kritzinger sequence with the Fibonacci sequence (left) and the Ostromoukhov sequence (right). Values are calculated every 1\,000 points and scaled by $n/\log(n)$. The black line corresponds to the best theoretical upper bound on the asymptotic discrepancy constant by Ostromoukhov. We clearly notice that this asymptotic constant is much better than the discrepancy of the Ostromoukhov sequence for a million points. While the asymptotic discrepancy order of the Ostromoukhov sequence is the best known to this day, it is outperformed by the Kritzinger sequence for the first million points.}
    \label{fig:res_1M}
\end{figure}

We present in this paper a simple algorithm to compute the elements of the Kritzinger sequence much faster. With this algorithm, we are able to compute the elements of the sequence up to millions of points. Comparing the resulting discrepancy values with both the Kronecker sequence with golden ratio and the current theoretical best sequence by Ostromoukhov in~\Cref{fig:res_1M} shows that the Kritzinger sequence outperforms both of them. Not only is it more regular than previous sequences, but it is also very robust relative to the starting points. As shown in~\Cref{sec:rob}, even when initializing the sequence with arbitrarily bad points, the sequence naturally corrects itself extremely quickly, with no noticeable impact long-term. It is especially interesting that a formulation based on the $L_2$ star discrepancy (see~\Cref{sec:L_2}) is able to give such good results for the $L_{\infty}$ star discrepancy. Finally, we provide methods to compute the natural extension of the sequence to dimensions 2 and 3 in~\Cref{sec:High}, showing that it stays competitive with well-known low-discrepancy sequences such as the Sobol' sequence.

\section{Competitiveness of the Kritzinger Sequence in Dimension One}

\subsection{Generating More Points: from Thousands to Millions}\label{sec:1d}

At each step, we want to compute efficiently the point $y$ minimizing the following function

$$F(y,P)= (n+1)y^2-y-2\sum_{x \in P}\max(x,y).$$

We build the sequence by adding at each step $\argmin_{y \in [0,1)}F(y,P)$, where $P$ is the set we have obtained so far. The essential property we will rely on to efficiently compute the sequence is based on the following theorem by Kritzinger~\cite{Kritz}.

\begin{theorem}\cite[Theorem 2]{Kritz} Let $P^*:=(x_n)_{n \in \mathbb{N}}$ be the sequence generated by the above method. Let $\Gamma_{n}:=\{(2i+1)/(2(n+1)):i\in \{0,\ldots,n\}\}$. Then, for all $n \in \mathbb{N}$, we have $x_n \in \Gamma_{n}$ and $x_n$ is different from all previous $x_k$, $k<n$.
\end{theorem}

Consider that our set is currently fixed with $n$ points $x_1<\ldots<x_n$ (the ordering does not correspond to the order they were added in). The key argument behind this theorem is to see that on each interval $[x_i,x_{i+1}]$, where we set $x_0=0$ and $x_{n+1}=1$, $F(y,P)$ is a second-order polynomial in $y$ with positive leading coefficient. This is because all the maxima can be replaced either by a point coordinate or $y$. It therefore has only a single minimum on each interval, and the values of $\Gamma_p$ naturally come from the coefficients of the polynomial. One can naturally define $F_i(y,P):=(n+1)y^2-A_iy-B_i$ to be the polynomial $F(y,P)$ obtained on the interval $[x_i,x_{i+1}]$.

By the above theorem, we only have to check $n+1$ possible values to find the next point to add to our sequence. This can also be done by finding the minimum of the $(n+1)$ $F_i(y,P)$ polynomials and keeping the smallest. While the first method requires recomputing $F(y,P)$ for every element, in linear time, it is possible to infer $F_{i-1}(y,P)$ from $F_i(y,P)$ in the second. Indeed, we know that as we move from one interval to the next, exactly one maximum $\max(y,x_i)$ in the function $F(y,P)$ will shift from being equal to $y$ to equal to $x_i$ (or vice versa). For the polynomial coefficients, this translates to $A_{i-1}=A_i-2$ and $B_{i-1}=B_i-2x_i$. We can find the minimum of each polynomial in constant time as it is a degree two polynomial, and obtain the next polynomial in constant time. Since we know that $F_{n}(y,P)=(n+1)y^2-(2n+1)y$, we can find the global minimum in linear time in the number of points already in the sequence.
Overall, this translates to a complexity of $O(N^2)$ to compute the first $N$ elements of the sequence. Finally, we remark that keeping the points sorted can be done in constant time as we build the sequence element by element. 

This is theoretically faster than Remark 2 from \cite{Kritz}, which says that if an interval $[\ell/n,(\ell+1)/n]$ already contains a point then it will not contain the next point. Indeed, checking if each interval already contains a point takes constant time, with an extra linear time to compute the local solution. If there are multiple empty intervals, this will take more time. We can combine the two together, but the gains are negligible. 

\begin{figure}[h]
    \centering
    \includegraphics[width=0.45\textwidth]{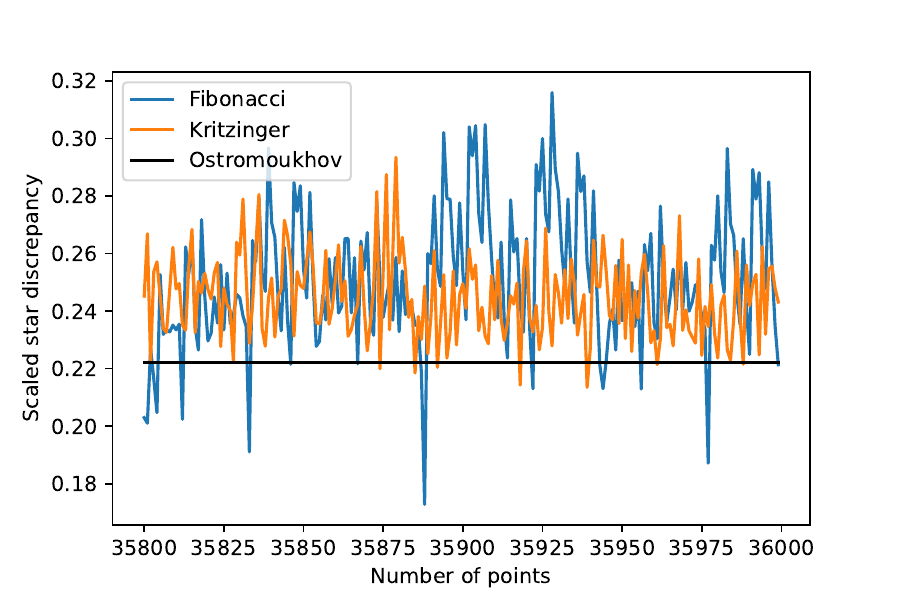}
    \hfill
    \includegraphics[width=0.45\textwidth]{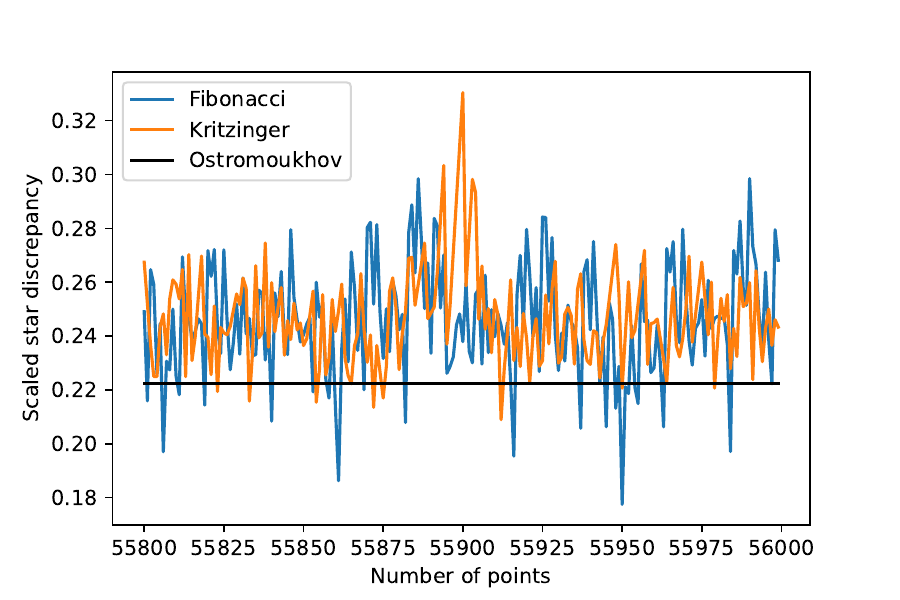}
    \caption{Localized plots in regions where the Kritzinger sequence was not performing well. Discrepancy values are calculated for all $n$ and scaled by $n/\log(n)$.}
    \label{fig:1dloc}
\end{figure}

\begin{figure}[h]
    \centering
    \includegraphics[width=0.5\textwidth]{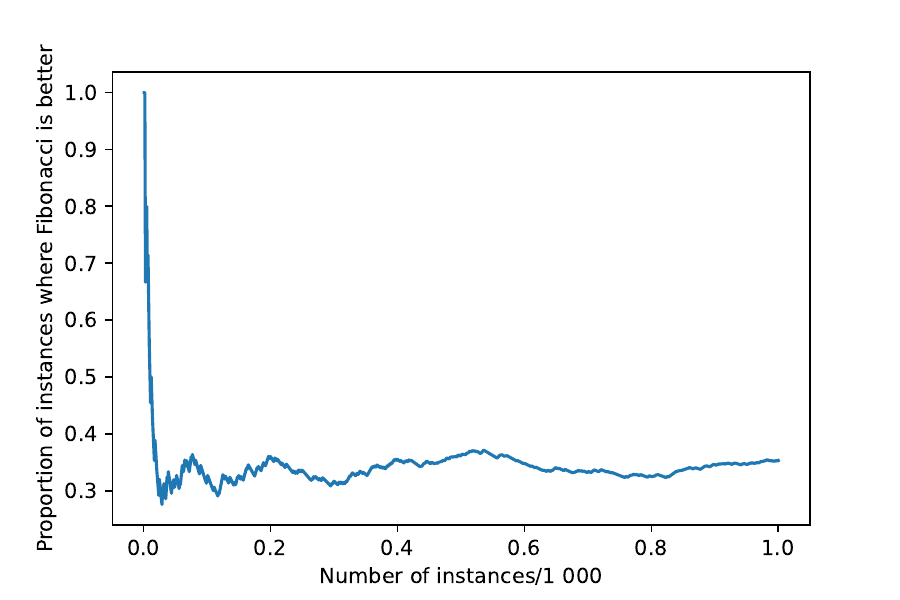}
    \caption{Proportion of $n$ for which the Fibonacci sequence is better than the Kritzinger one, initialized in $x=0.5$, counting only one instance per 1000 points.}
    \label{fig:compa}
\end{figure}

With a discrepancu value computed every 1\,000 points, \Cref{fig:res_1M} compares the first million points of the Kritzinger sequence with the first million of the Kronecker sequence with golden ratio (also known as Fibonacci sequence) and the Ostromoukhov sequence which we introduced in \Cref{sec:intro}. \Cref{fig:res_1M} shows that the Kritzinger sequence is competitive with, and even on average better than, the golden ratio Kronecker sequence for the first 1\,000\,000 points. Plots with one-point steps in~\Cref{fig:1dloc} show that the spikes in the Kritzinger sequence's discrepancy values are extremely localized. We note that the big spikes represent only \emph{one or two} successive sub-par point choices out of thousands of points. \Cref{fig:compa} describes the proportion of instances for which the golden ratio Kronecker sequence is better than the Kritzinger sequence. These experiments show the Kritzinger sequence is reliably outperforming the Fibonacci sequence, being better for around 65\% of the instances. This empirical convergence provides further evidence for the following conjecture by Kritzinger.

\begin{conjecture}\cite[Conjecture 2]{Kritz}
    The Kritzinger sequence has an $L_{\infty}$ star discrepancy in $O(\log(n)/n)$ in one dimension.
\end{conjecture}

\subsection{Robustness of the Kritzinger sequence}\label{sec:rob}

\begin{figure}[h]
    \centering
    \includegraphics[width=0.5\textwidth]{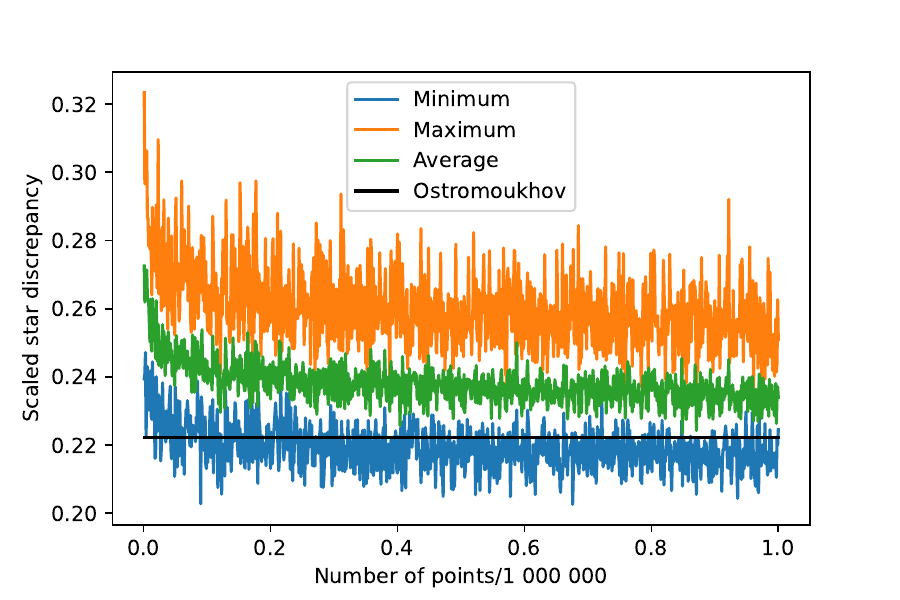}
    \caption{Best, average and worst scaled discrepancy for 11 sequences with different single starting points for the Kritzinger sequence in dimension 1.}
    \label{fig:1d1init}
\end{figure}

\textbf{Impact of the initial point:} In all our work so far, the sequence was naïvely initialized with a single point in 0.5. We performed experiments with $x_0=0.1j$, for $j=0$ to $9$ and an extra $x_{10}=0.9999$. All the sequences performed similarly as~\Cref{fig:1d1init} shows. For readibility, we plot the minimum, maximum, and average scaled discrepancy for each of these sequences. All of these have similar behaviors and appear to have a discrepancy in $O(\log(n)/n)$, close to the best known constant.

\begin{figure}
    \centering
    \includegraphics[width=0.5\textwidth]{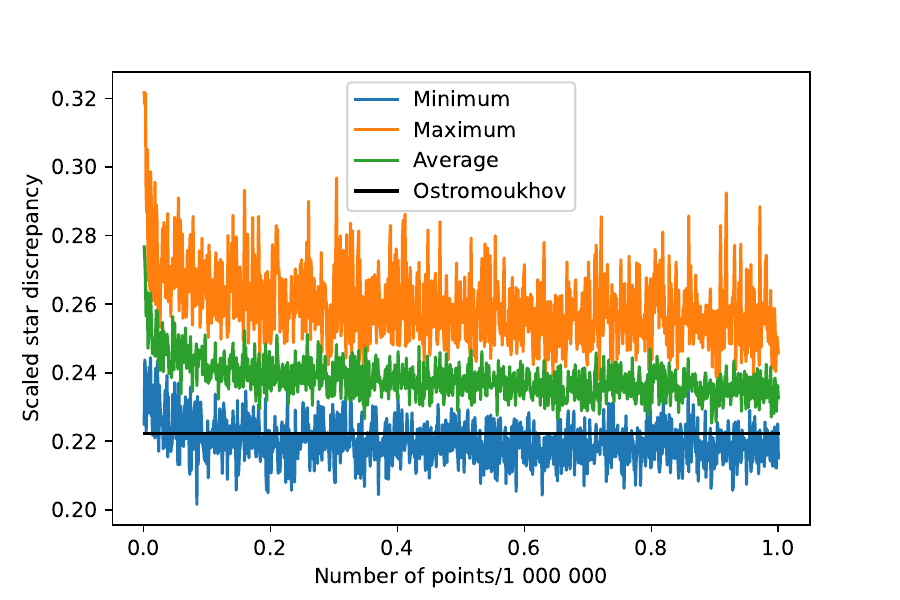}
    \caption{Best, average and worst scaled discrepancies for 6 different sets with 5 random initial points in dimension 1.}
    \label{fig:1d5init}
\end{figure}

Changing this single initialization point to 5 randomly selected values in~\Cref{fig:1d5init} has also no noticeable impact on the quality of the obtained sequence. We note that both in this case and the previous single point initialization, there is no meaningful difference between the different runs. All seem to have the same behavior, there is not one that performs better more often or another that is more often worse.

\begin{figure}[h]
    \centering
    \includegraphics[width=0.6\textwidth]{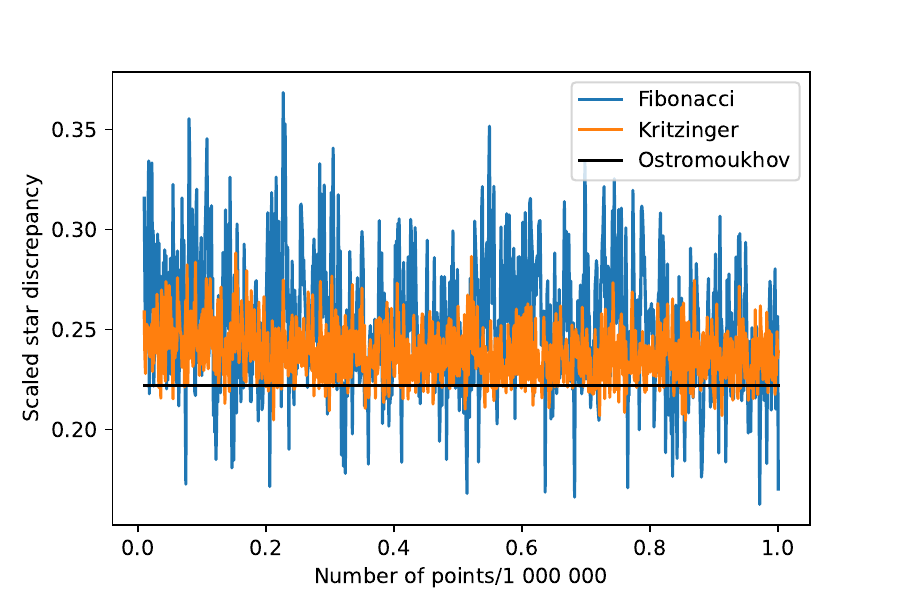}
    \caption{Scaled discrepancy values with a bad initial set of 100 points, for $n=10\,000$ to one million.}
    \label{fig:bad}
\end{figure}

\textbf{Correcting a bad initial set:} Finally, we initialize the sequence with a large set of badly chosen points: 100 points from 0 to 0.01 with a step-size of $10^{-4}$. While the discrepancy values are unsurprisingly relatively poor initially, they are as good as those obtained with a single initial point at the 10\,000 point mark, and continue to perform just as well as the previous sequences afterwards, as~\Cref{fig:bad} shows. The initial values are not included in the plot to make it readable. The first 9 values (for $n=1000$ to $n=9000$) nevertheless come very close to the discrepancy that would be obtained with only the initial points in the box $[0,0.01)$. For example, for 7000 points, we obtain a discrepancy of 0.00438, whereas a set of 7\,000 points with 100 points in $[0,0.01]$ has a discrepancy of at least 0.00428.

These experiments suggest the relatively simple greedy approach introduced by Kritzinger can be used not only to build a low-discrepancy sequence, but also to correct point sets, which is not something either of the two constructions described in~\Cref{sec:intro} can do.

\section{Extending the Sequence to Dimensions 2 and 3}\label{sec:High}

\subsection{The \texorpdfstring{$L_2$}{L_2} Star Discrepancy}\label{sec:L2}
We have not mentioned so far the origin of this functional. The $L_2$ star discrepancy is a common discrepancy measure, used in particular because it is much easier to compute than the $L_{\infty}$ star discrepancy. Indeed, it can be computed in quadratic time by Warnock's formula~\cite{Warnock} given below, whereas the $L_{\infty}$ star discrepancy computation is NP-complete~\cite{complexity}, with the best algorithm known to this day in $O(n^{1+d/2})$~\cite{DEM}, where $d$ is the dimension of the point set.

\begin{equation}\label{eq:Warnock}
d^{*}_2(P)=\frac{1}{3^d}-\frac{2^{1-d}}{n}\sum_{i=1}^n\prod_{k=1}^{d} (1-x_{i,k}^2)+\frac{1}{n^2}\sum_{i,j=1}^n\prod_{k=1}^{d} (1-\max(x_{i,k},x_{j,k})),
\end{equation}
where $x_{i,k}$ is the $k$-th coordinate of the $i$-th point of $P$. The functional defined by Kritzinger should be seen as the contribution of a \emph{single} point in equation~\eqref{eq:Warnock} in one dimension. Generalizing the functional to higher dimensions can be done naturally by considering the contribution of a single point in equation~\eqref{eq:Warnock} in the desired dimension.

In dimension $d$, when adding the $(n+1)$-th point, this corresponds to
\begin{equation}
    F_d(y,P):=2^{1-d}(n+1)\prod_{k=1}^{d} (1-y_k^2)+\prod_{k=1}^d (1-y_k) +2\sum_{i=1}^{n}\prod_{k=1}^d (1-\max(x_{i,k},y_k))
\end{equation}

One immediately sees that even without considering the maxima, we no longer have a second-order polynomial in $y$. For example in dimension 2, we now have a polynomial in $y_1$ and $y_2$, with the highest order term in $y_1^2y_2^2$. There are two main issues with this. The first is that this is not easy to solve exactly unlike in the one-dimensional case. %While there is software such as msolve~\cite{berthomieu}, it is too slow for our purposes. 
The second is that to decompose $[0,1)^d$ in such a way that all the maxima can be removed as in~\Cref{sec:1d} requires $n^d$ boxes, for each of which we would need to find the minimal argument of a non-trivial polynomial.

\subsection{Construction methods}
We propose different methods to compute $\argmin_{y \in [0,1)^d}F_d(y,P)$. While they can be generalized to any dimension, we will only focus on dimensions 2 and 3 where they remain tractable and provide relatively accurate results. Our first approach is via Non-Linear Programming. This \emph{exact} method is inspired by previous work in~\cite{CDKP} to compute optimal $L_{\infty}$ star discrepancy sets. The idea is to reformulate our problem with an objective (the function we want to minimize), constraints and variables (some of which correspond to the points' coordinates). In this case, our objective will be the minimization of $F_d(y,P)$ and the main variables correspond to the points' coordinates. A detailed description of this method is provided in~\Cref{app:NLP}.

\begin{figure}[h]
    \centering
    \includegraphics[width=0.45\textwidth]{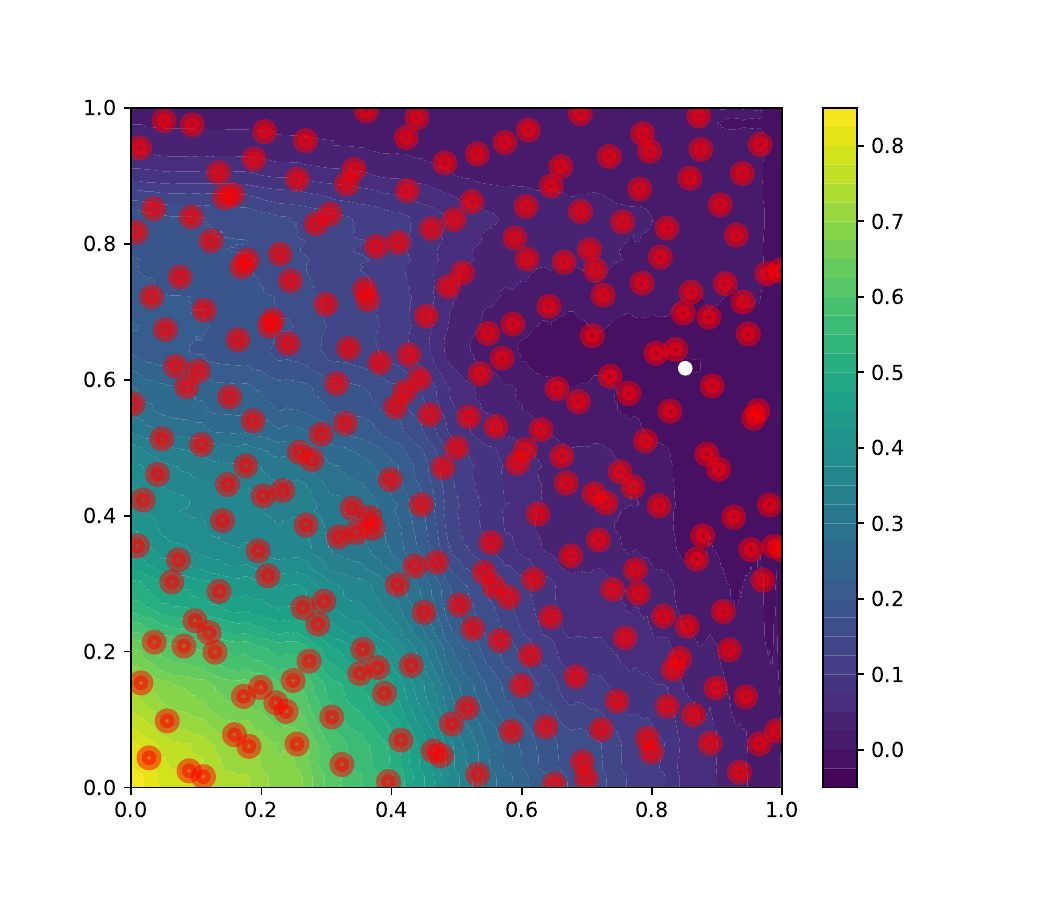}
    \includegraphics[width=0.45\textwidth]{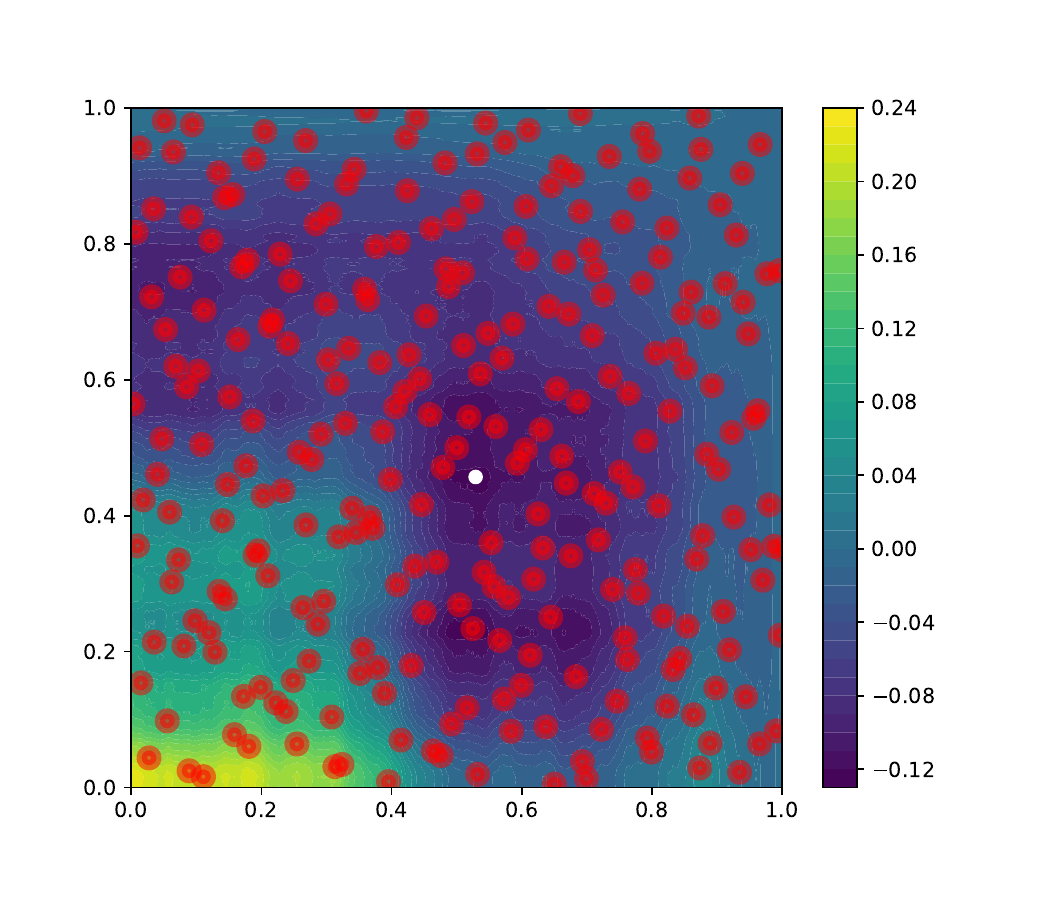}
    \caption{Plots of $F(y,P)$ for the Kritzinger sequence initialized in $(0.5,0.5)$ with 247 (left) and 266 points (right). The point minimizing $F(y,P)$ is shown in white (respectively the 248-th and 267-th points of the sequence). $F(y,P)$ is very smooth, suggesting gradient descent methods can be used locally.}
    \label{fig:SobL2}
\end{figure}

We also use a diverse set of heuristic approaches. These include some naïve search methods to find the local minimum, such as random search or a grid-based search. Quasi-Monte Carlo points could also form the basis for the exploration. In addition, the function appears to be very smooth as~\Cref{fig:SobL2} shows. We can therefore expect gradient descent to work relatively successfully. One does have to be careful the function $F(y,P)$ is not differentiable everywhere, as such we have to perform a gradient descent for each $F_{i,j}(y,P)$, and not just one general one, and keep the global minimum.

\subsection{Results in dimensions 2 and 3}

In this section, we present our results in dimensions 2 and 3. While the methods are valid for any dimension, they get more expensive as the dimension increases. Furthermore, there is a great instability in the point choices. In $1d$, picking a point slightly off would have little impact for the next few points, as we know that they respect a specific structure. This is no longer the case in higher dimensions: choosing one point very slightly differently can lead to instant changes for the next point and then lead to very different sequences. We initialized 10 different sequences with $(0.5,0.5)$ and used the random heuristic on them separately to obtain 500 points. While they all had very similar discrepancy values over the 500 points, the sequences generally became very different around the 70 points mark. In some cases, two of the sequences were identical up to 70 points and did not have a single point in common after the 80th. As the dimension increases, this leads to much greater mistakes in the sequence.

\begin{figure}[h]
    \centering
    \includegraphics[width=0.47\textwidth]{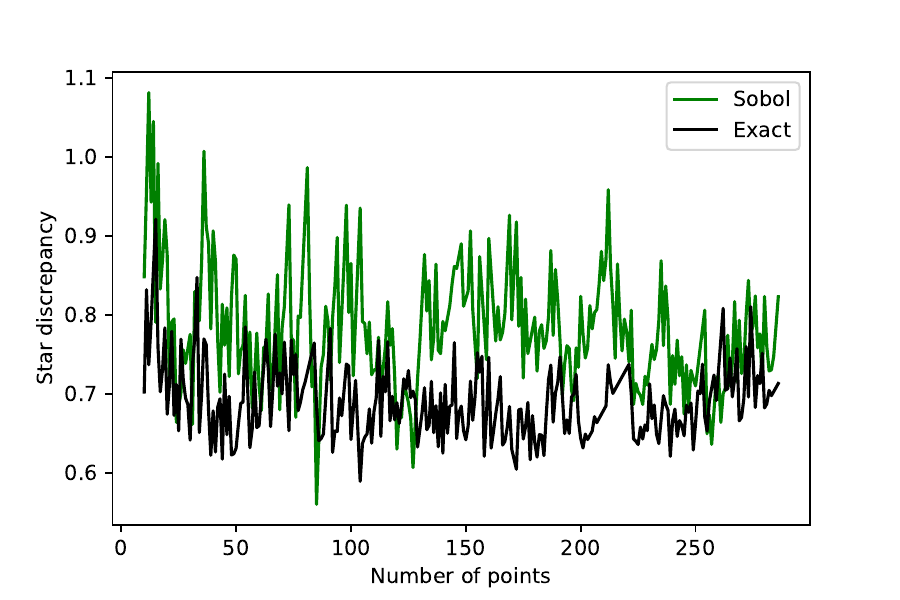}
    %\caption{Kritzinger sequence initialized in (0.5,0.5) compared with the Sobol' sequence, discrepancies scaled by $n/\log(n)$.}
    \includegraphics[width=0.47\textwidth]{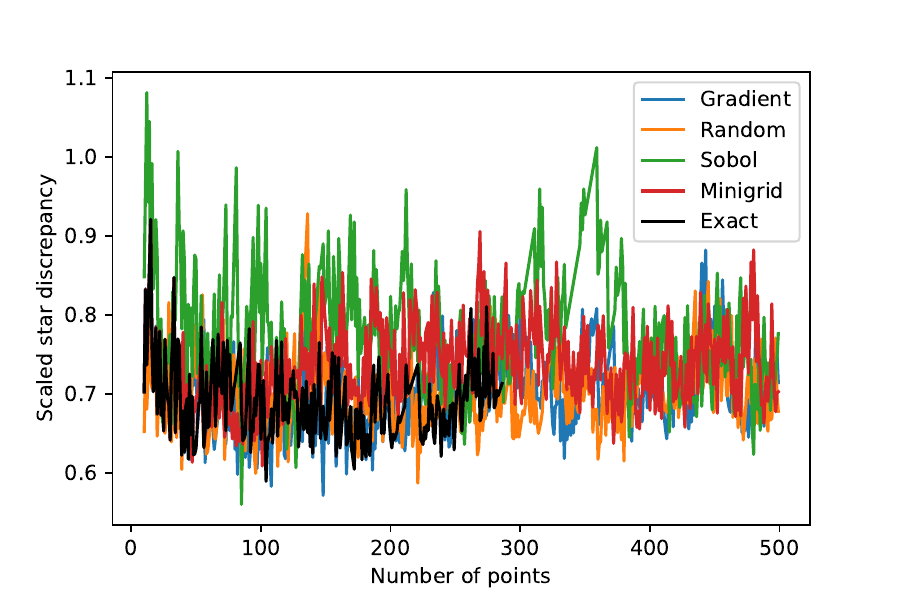}
    %\caption{Comparison of the different methods initialized in (0.5,0.5) with the Sobol' sequence, discrepancies scaled by $n/\log(n)$.}
    \caption{On the left, comparison of the exact Kritzinger sequence initialized in (0.5,0.5) compared with the Sobol' sequence. On the right, comparison of the different methods initialized in (0.5,0.5) with the Sobol' sequence. In both, discrepancies are scaled by $n/\log(n)$.}
    \label{fig:ex2d}
\end{figure}

\begin{figure}[h]
    \centering
    \includegraphics[width=0.5\textwidth]{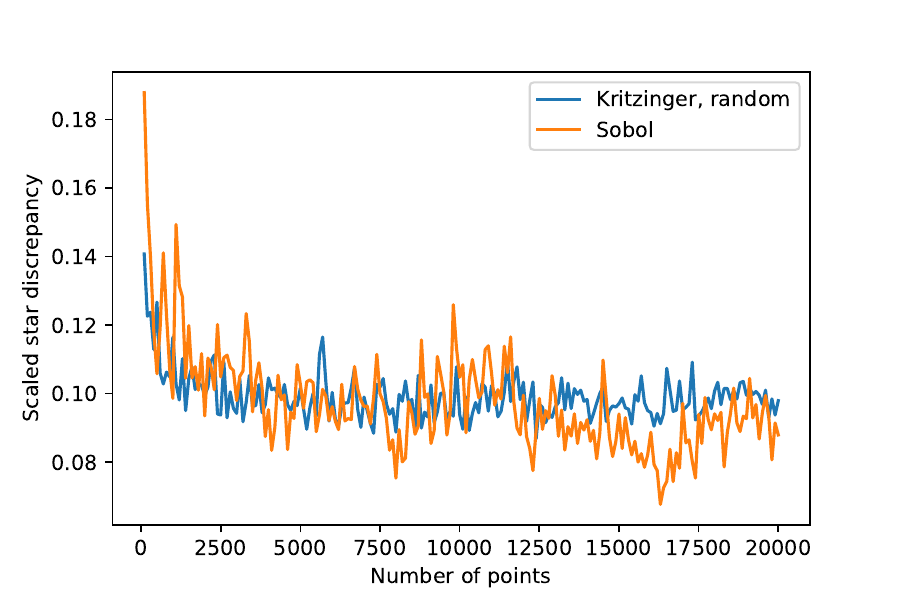}
    \caption{Random heuristic and Sobol' sequence $L_{\infty}$ star discrepancies for up to 20 000 points in dimension~2. $L_{\infty}$ star discrepancy values are scaled by $n/\log^2(n)$. }
    \label{fig:20K2d}
\end{figure}

Nevertheless, our results show that the quality of the sequence remains robust, at least in 2 dimensions. In particular, \Cref{fig:ex2d} (left) gives the $L_{\infty}$ star discrepancy of the first 277 points of the Kritzinger sequence initialized in $(0.5,0.5)$, compared to the Sobol' sequence. This is scaled by $n/\log(n)$ (not squared, otherwise the plot would have needed to be truncated for readability). \Cref{fig:ex2d} (right) then compares the approximate methods with the exact sequence and the Sobol' sequence in the same context. We see that our approximate methods produce sequences that are quite similar to the exact sequence, at least for the discrepancy values obtained. All of them are competitive with the Sobol' sequence for all values shown. This suggests to some degree that as in the one-dimensional experiments with different starts, greedy $L_2$ sequence construction is not too dependent on the exact points themselves for good performance. Even when increasing $n$ further as in~\Cref{fig:20K2d}, we observe that the random method generates a sequence whose $L_{\infty}$ star discrepancy values are comparable to those of the Sobol' sequence over the first 13\,000 points. While there seems to be a big difference around 16\,000 points, one should remember that the construction method of the Sobol' sequence leads to better values when $n=2^k$ for $k \in \mathbb{N}$. Combining this with the low precision of the method partly explains the discrepancy gap. 

\begin{figure}[h]
    \centering
    \includegraphics[width=0.5\textwidth]{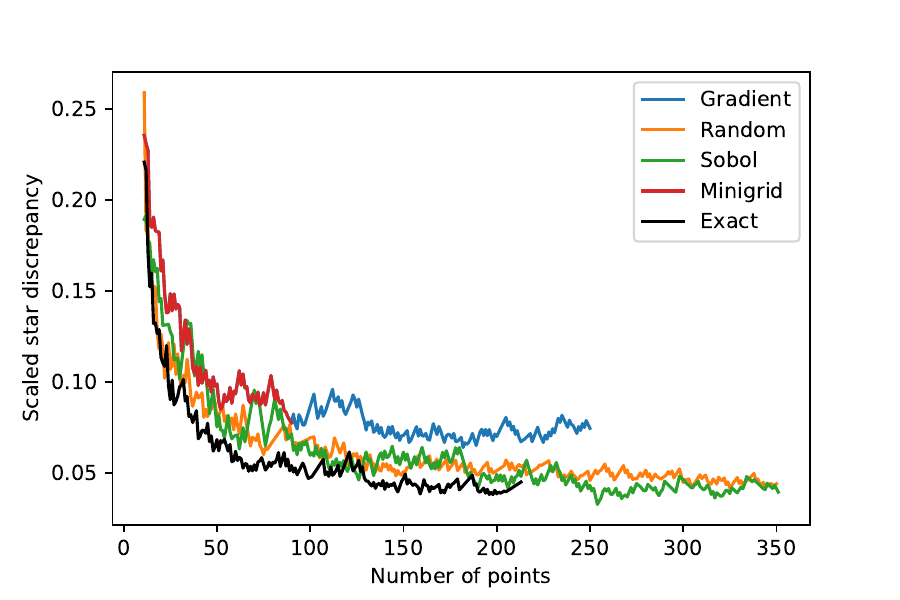}
    \caption{Comparison of our different methods in dimension 3 with the Sobol' sequence. Values are scaled by $n/\log^3(n)$. %Right: Comparison of our approximate methods (smaller number of samples for random than in the left plot) with the Halton and Sobol' sequences.
    }
    \label{fig:ex3d}
\end{figure}

Finally, \Cref{fig:ex3d} describes our results in dimension 3. The exact Kritzinger sequence consistently outperforms the Sobol' sequence, while the random approach seems to perform decently. Both the gradient and discrete grid approaches are a lot less successful. We note that while we are able to compute more points with the random approach, their quality degrades around $n=500$. \Cref{fig:ex3d} shows that our random points approach is generally outperformed by the Sobol' sequence. Nevertheless, the discrepancy order seems to still be $O(\log(n)^3/n)$, despite the imprecisions in the construction.

Our empirical results in 2 and 3 dimensions, suggesting quite strongly that the discrepancy is of the same order as for the Sobol' sequence, lead us to formulate the following conjecture.
\begin{conjecture}
    The Kritzinger sequence has a discrepancy of $O(\log^d(n)/n)$ in dimension $d$.
\end{conjecture}

\textbf{Acknowledgments:} The author would like to thank Stefan Steinerberger and Carola Doerr for their helpful comments, ideas and feedback, in particular for the suggestion of testing the robustness relative to different initialization points. We gratefully acknowledge the support of the Leibniz Center for Informatics,
where several discussions about this research were held during the Dagstuhl Seminar
``Algorithms and Complexity for Continuous Problems'' (Seminar ID 23351).

\appendix
\section{Non-Linear Programming Formulation}\label{app:NLP}
 We can express our problem of optimizing the placement of a point in an existing set with regards to the $L_2$ star discrepancy as a non-linear mathematical programming problem.

\begin{subequations}\label{LP model}
\begin{align}
    \min \;\;&-\frac{n+1}{2}v_3+(1-y_1)(1-y_2)+2\sum_{j=1}^nt_ju_j&&\nonumber \\
    &x_{2j-1}=x_{j,1} && \forall j=1,\ldots,n \label{eq:fixx}\\
    &x_{2j}=x_{j,2} && \forall j=1,\ldots,n \label{eq:fixy}\\
    &r_j-1 \leq x_{2j-1}-y_1 && \forall j=1,\ldots,n \label{eq:5c}\\
    &r_j \geq x_{2j-1}-y_1 && \forall j=1,\ldots,n\\
    &s_j-1 \leq x_{2j}-y_2 && \forall j=1,\ldots,n\\
    &s_j \geq x_{2j}-y_2 && \forall j=1,\ldots,n \label{eq:5f}\\
    & t_j=(1-r_j)y_1+r_jx_{2j+1} && \forall j=1,\ldots,n \label{eq:5g}\\
    & u_j=(1-s_j)y_2+s_jx_{2j+2} && \forall j=1,\ldots,n \label{eq:5h}\\
    &v_1=y_1y_1 &&\label{eq:5i}\\
    &v_2=y_2y_2 &&\\
    &v_3=(1-v_1)(1-v_2) &&\label{eq:5k}\\
    & r_{j},s_{j} \in \{0,1\},&& \forall j=1,\dots,n, \nonumber \\
    &t_{j},u_{j}\in[0,1]\,&& \forall j=1,\dots,n, \nonumber \\
    &x_{j}\in[0,1]\, \forall j=1,\dots,2n,\; y_1,y_2,v_1,v_2,v_3 \in [0,1] &&\nonumber
\end{align}
\end{subequations}

Model~\eqref{LP model} describes the objective and constraints used in the two-dimensional case, this can be easily generalized to higher dimensions. The objective corresponds to the minimization of $F((y_1,y_2),P)$, with some reformulations explained below. The constraints~\eqref{eq:fixx} and~\eqref{eq:fixy} fix the values of the previous points. Each of the maximum terms in the final sum needs to be reformulated for the solver. For this, we introduce binary variables in equations~\eqref{eq:5c} to~\eqref{eq:5f} (the $r_j$ for the first dimension and $s_j$ for the second) that are equal to 1 if the constant is greater, and 0 if our variable is bigger. For each of these binary variables, two constraints are required:
\begin{equation*}
    r_j-1 \leq x_{2j-1}-y_1\;\;\;\; \forall j=1,\ldots,n     
\end{equation*}
\begin{equation*}
    r_j \geq x_{2j-1}-y_1 \;\;\;\; \forall j=1,\ldots,n
\end{equation*}
The first constraint imposes that if the constant coordinate is smaller, then $r_j$ is equal to 0. The second fixes $r_j$ to 1 if the constant coordinate is bigger. Variable $t_j$ in constraint~\eqref{eq:5g} then plays the role of the maximum function for the first coordinate, while variable $s_j$ in constraint~\eqref{eq:5h} plays a similar role for the second coordinate. While no other constraints are theoretically required, solvers such as Gurobi \cite{Gurobi} cannot handle products of more than two variables easily. Terms such as $y_1^2y_2^2$ have to be replaced. For this, we introduce three extra variables $v_1$ to $v_3$ in constraints~\eqref{eq:5i} to~\eqref{eq:5k}, simply to reformulate the objective.

\bibliographystyle{alpha}
\bibliography{refs}

\end{document}